\documentclass{jfm}
\usepackage{graphicx}
\usepackage{epstopdf, epsfig, mathrsfs, color}
\usepackage{hyperref}

\shorttitle{Universality of capillary rising in corners}
\shortauthor{J. Zhou and M. Doi}

\title{Universality of Capillary Rising in Corners}

\author{Jiajia Zhou\aff{1,2}
  \corresp{\email{jjzhou@buaa.edu.cn}}
  \and Masao Doi\aff{2}
  \corresp{\email{masao.doi@buaa.edu.cn}}}

\affiliation{\aff{1}Key Laboratory of Bio-Inspired Smart Interfacial Science and Technology of Ministry of Education, School of Chemistry, Beihang University, Beijing 100191, China
\aff{2}Center of Soft Matter Physics and its Applications, Beihang University, Beijing 100191, China}

\begin{document}

\maketitle

\begin{abstract}
We study the dynamics of viscous capillary rising in small corners between two curved walls described by a function $y=cx^n$ with $n \ge 1$.
Using Onsager principle, We derive a partial differential equation that describes the time evolution of the meniscus profile. 
By solving the equation both numerically and analytically, we show that the capillary rising dynamics is quite universal.
Our theory explains the surprising finding by Ponomarenko \emph{et al.} that the time dependence of the height not only obeys the universal power law of $t^{1/3}$, but also that the prefactor is almost independent of $n$.
\end{abstract}


\section{Introduction}
\label{sec:introduction}

It is well-known that when a capillary tube is brought in contact with a wetting fluid, the fluid rises in the tube and eventually reaches the Jurin's height \citep{dBQ}.  
A related setup is when a cornered geometry consisting of two intersecting plates is brought in contact with the fluid. 
In this case, a finger-like fluid quickly forms at the corner, and the tip of the finger keeps rising indefinitely (since the equilibrium position of the tip is infinitely high).
The earliest study of the capillary rising in corners can be dated back to the 18th century: Taylor conducted experiments on the fluid rising in a small-angle corner formed by two nearly parallel plates. 
He identified that the equilibrium shape of the meniscus is a hyperbola \citep{Taylor1710}. 
In the paper right after Taylor's, \citet{Hauksbee1710} confirmed and quantified Taylor's observation.
Other reports on the equilibrium meniscus can be found in works of \citet{Langbein1990} and \citet{Finn1999, Finn2002a}.

Even though the equilibrium theory of the meniscus in a cornered geometry is well-established, the understanding on the dynamics was quite recent. 
The time evolution of the meniscus is governed by several factors.
The driving force is the capillary force which tends to minimize the interfacial energy, and the wetting fluid tends to maximize its coverage on the solid surfaces. 
The rising of the fluid is hindered by the viscous friction and the gravity.
In situations when the gravity can be ignored, the propagation of the meniscus front obeys the classical Lucas-Washburn's $t^{1/2}$ scaling \citep{Lucas1918, Washburn1921, Dong1995, Weislogel1998}.
When gravity is considered, the tip of the meniscus rises with a different $t^{1/3}$ scaling.
This result was first derived by \citet{Tang1994}.
\citet{Higuera2008} developed a more complete theory for the case of two flat plates forming a small angle.
They derived a partial differential equation for the time evolution of the meniscus shape based on the lubrication approximation, and derived the $t^{1/3}$ scaling law from this equation.
\citet{Ponomarenko2011} conducted experiments of capillary rising in corners of different geometries where walls are curved and are described by functions $y=c x^n$ with $n \ge 1$. 
They found that the meniscus front obeys the same $t^{1/3}$ scaling as the flat wall. 
Corroborated with an accompanying scale analysis, they have established that the position of the meniscus front $Z_m$ obeys the following equation
\begin{equation}
  \label{eqn:1} 
  \frac{Z_m}{a_c} = C \left( \frac{\gamma t}{\eta a_c} \right)^{1/3} \, .
\end{equation}
In equation (\ref{eqn:1}), the length and time are scaled by the capillary length $a_c=\sqrt{\gamma/ \rho g}$ and $\eta a_c/ \gamma$, respectively ($\gamma$, $\eta$, $\rho$ are surface tension, viscosity and density of the liquid and $g$ is the gravity constant), and $C$ is a numerical factor.
The experimental results of \citet{Ponomarenko2011} were in good agreement with equation (\ref{eqn:1}). 
Quite surprisingly, they also found the experimental data collapse to a universal curve, having the same numerical factor independent of $n$.  
This means that the dynamics of the meniscus rise is quite universal, independent of the shape of the corner.

In this paper, we study the dynamics of viscous capillary rising at a general corner. 
Using Onsager principle \citep{Onsager1931, Onsager1931a, DoiSoft}, we derive a partial differential equation that describes the time evolution of meniscus profile, and solve it both numerically and analytically.   
We show that 
(1) the advance of the meniscus front follows the time-scaling of $t^{1/3}$, and 
(2) the front factor $C$ changes only 10\% when $n$ changes from  1 to 5.   
This explains the universality found by \citet{Ponomarenko2011}.

\section{Capillary rising in a corner}
\label{sec:model}

\begin{figure}
  \centering
  \includegraphics[width=0.8\textwidth]{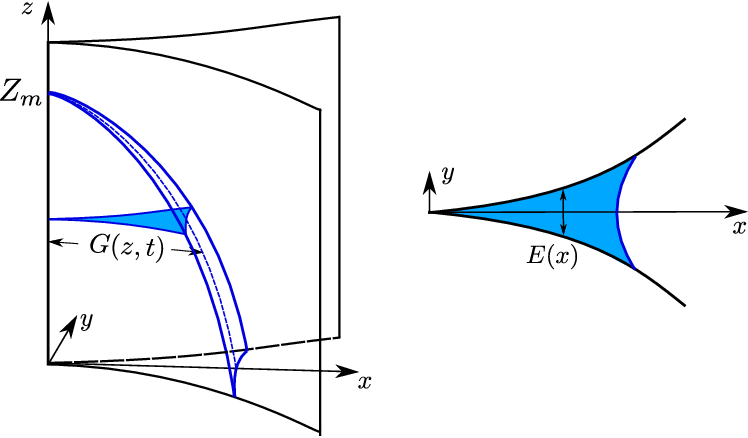}
  \caption{Schematic picture of the capillary rising in a power-law corner.}
  \label{fig:sketch}
\end{figure}

We consider the capillary rising in a corner formed by two surfaces as shown in figure~\ref{fig:sketch}.  
We take the coordinate system with $z$-axis along the intersection of the surfaces and 
$x$-axis bisecting the surfaces.
The two surfaces forming the corner are described by a function $y= \pm E(x)/2$.
The meniscus is described by the profile in the $x$-$z$ plane, given by the function $x=G(z,t)$. 
The bottom of the meniscus is located at $z=0$ and in contact with the fluid reservoir. 
The tip of the meniscus is denoted by $z=Z_m$. 

\subsection{Onsager principle}

To derive the time evolution equation for the meniscus profile $G(z,t)$, we use Onsager principle, 
the variational principle proposed by \citet{Onsager1931, Onsager1931a} for general irreversible processes. 
In the present context, this principle can be regarded as a variational formulation of 
Stokesian hydrodynamics for problems that have moving boundaries.  
In Stokesian hydrodynamics, the velocity field is determined by the minimum energy dissipation principle if the velocity at the boundary is known.  
Onsager principle here can be viewed as an extension of this principle to determine the motion of the boundary.

Our objective here is to determine the time evolution of $G(z,t)$. 
For this purpose,  we construct a functional called Rayleighian, which is a functional of $\dot G(z,t)$, the time derivative of $G(z,t)$.  
The Rayleighian $\mathscr{R}[\dot G(z,t)]$ is a sum of two terms
\begin{equation}
     \mathscr{R}[\dot G(z,t)] = \dot F [\dot G(z,t)] + \Phi [\dot G(z,t)] ,
\end{equation}
both being a functional of $\dot G(z,t)$.
The first term $\dot F [\dot G(z,t)]$ represents the change rate of the free energy when the boundary is moving at rate $\dot G(z,t)$.  
The dissipation function $\Phi [\dot G(z,t)]$ represents half of the energy dissipation rate (or the entropy
production rate) taking place in the system when the boundary is changing at rate $\dot G(z,t)$. 
Onsager principle states that $\dot G(z,t)$ is obtained by minimizing the functional  $\mathscr{R}[\dot G(z;t)]$.

In the following calculation, we do not consider $\dot G(z,t)$ explicitly. 
Rather, we take the volume flux $Q(z,t)$ of fluid flowing across the plane at $z$ as an independent variable and express the Rayleighian as a functional of $Q(z,t)$.
$\dot G(z,t)$ and $Q(z,t)$ are related to each other by the conservation equation.  
Let $A(G)$ be the area of the region $\{(x,y) \, | \, 0<x<G, |y|<E(x)/2 \}$, i.e., 
\begin{equation}
  \label{eq:AG}
  A(G)= \int_0^G E(x) \, \mathrm{d} x . 
\end{equation}
Then the conservation equation for the fluid volume is written as
\begin{equation}
  \label{eq:cons}
  \frac{\partial A}{\partial t} = A' \dot{G} = - \frac{\partial Q}{\partial z},
\end{equation}
where the prime denotes the derivative with respect to $G$, $A'=\partial A / \partial G$.

In the following we shall first calculate $\dot F $ and $\Phi$ expressed as a functional of $Q(z,t)$ and determine the flux $Q(z,t)$ by minimizing the Rayleighian. 
The time evolution equation for $G(z,t)$ is given by the conservation equation (\ref{eq:cons}).

\subsection{Free energy}

The free energy of the system is given by
\begin{equation}
  \label{eq:freeE}
  F[G(z,t)] = \int_0^{Z_m} \Big( \rho g A(G(z,t)) z - 2 L(G(z,t)) \gamma \cos\theta \Big)  \, \mathrm{d} z \,.
\end{equation}
The first term is the gravitational energy and the second term is the interfacial energy.
In equation (\ref{eq:freeE}) $L(G)$ is the contour length of the curve $y=E(x)/2$ for $0<x<G$
\begin{equation}
  L(G) = \int_0^G  \sqrt{1+ \frac{1}{4} 
    \left( \frac{\mathrm{d} E}{\mathrm{d} x} \right)^2} \,  \mathrm{d} x \, .
\end{equation}
$\theta$ is the equilibrium contact angle of the fluid on the solid surfaces.
We focus on the fluid that wets the surface, i.e., the contact angle $\theta$ is close to zero.
In writing the free energy in the form of (\ref{eq:freeE}), we have neglected the surface energy of the free surface.
In general, this contribution is of the order of $E(G)\gamma$.  
When the two surfaces are close to each other and for the fully-wetting fluid, we have $L(G) \sim G \gg E(G)$, thus the free surface contribution can be ignored.

The change rate of the free energy is
\begin{equation}
  \label{eq:Fdot}
  \dot{F} = \int_0^{Z_m} \Big( \rho g A' z - 2 L' \gamma \cos\theta \Big) \dot{G}  \, \mathrm{d} z 
  = \int_0^{Z_m} \Big( \rho g - 2 \frac{\partial (L'/A')}{\partial z} \gamma \cos\theta \Big) Q  \, \mathrm{d} z \, , 
\end{equation}
where we have used the conservation equation (\ref{eq:cons}) and the integration by part.

The equilibrium profile of the meniscus is determined by setting the integrand in (\ref{eq:Fdot}) to be zero
\begin{equation}
  \label{eq:eq}
  \frac{ \partial (L'/A') }{\partial z} = \frac{\rho g}{ 2\gamma \cos\theta} .
\end{equation}

\subsection{Dissipation function}

The dissipation function is calculated by the lubrication approximation.
In this approximation, the pressure $p$ is assumed to be constant in $x$-$y$ plane (i.e., $p$ depends on $z$ only), and the velocity has the $z$ component $v_z$ only. 
This requires the characteristic length scales in $x$- and $y$-directions are much smaller than that in the $z$-direction (see section \ref{sec:discussion} for more detailed discussion). 
The velocity $v_z$ is determined by the Stokes equation
\begin{equation}
  \eta \left( \frac{\partial^2}{\partial x^2} +   \frac{\partial^2}{\partial y^2} \right ) v_z = - \frac{\partial p}{\partial z} - \rho g ,
\end{equation}
with the boundary condition $v_z=0$ at the boundary $y= \pm E(x)/2$.  
The equation can be solved analytically by utilizing the fact that the length scale in $y$-direction, $E(x)$, is much smaller than the length scale in $x$-direction, $G(z)$.  
Hence the flow profile is essentially parabolic, and $v_z$ can be written as
\begin{equation}
   v_z(x,y,z,t)= \frac{3}{2} \, \bar{v}_z(x,z,t)
   \left[ 1 - \left( \frac{2y}{E(x)} \right)^2 \right] ,     
\end{equation}
where $\bar{v}_z(x,z,t)$ is the $y$-averaged velocity. 
In channel flow,  $\bar{v}_z(x,z,t)$ is proportional to $E^2(x)$, and therefore it can be written as
\begin{equation}
  \label{eq:vz}
  \bar{v}_z (x,z,t) = C(z,t) E^2(x).
\end{equation}
The flux is an area integration of the local velocity
\begin{equation}
  \label{eq:Qz}
  Q(z) = \int_0^{G(z)} E(x) \bar{v}_z(x,z,t)  \, \mathrm{d} x
  = \int_0^{G(z)} E^3(x) C(z,t)  \, \mathrm{d} x 
  = B(G) C(z,t) ,
\end{equation}
where the function $B(G)$ is given by 
\begin{equation}
  B(G) = \int_0^G E^3(x)  \, \mathrm{d} x \, . 
\end{equation}

The velocity is then given by equations (\ref{eq:vz}) and (\ref{eq:Qz})
\begin{equation}
  \bar{v}_z(x,z,t) = \frac{Q(z)}{B} E^2(x).
\end{equation}
The dissipation function is 
\begin{equation}
  \label{eq:Phi}
  \Phi = \frac{1}{2} \int_0^{Z_m} \int_0^{G(z)} 
  \frac{12 \eta}{E(x)} \bar{v}_z^2(x,z,t)  \, \mathrm{d} z \mathrm{d} x
  = \frac{1}{2} \int_0^{Z_m} \frac{12 \eta }{B} Q^2(z)  \, \mathrm{d} z \,.
\end{equation}

\subsection{Time evolution equation}

Given the change rate of the free energy (\ref{eq:Fdot}) and the dissipation function (\ref{eq:Phi}), the Rayleighian is obtained as
\begin{equation}
  \label{eq:Ray}
  \mathscr{R} = \dot{F} + \Phi 
  = \int_0^{Z_m} \Big( \rho g - 2 \frac{\partial (L'/A')}{\partial z} \gamma \cos\theta \Big) Q  \, \mathrm{d} z
  + \frac{1}{2} \int_0^{Z_m} \frac{12 \eta}{B} Q^2  \, \mathrm{d} z \, .
\end{equation}
The time evolution equation is derived from Onsager variational principle, $\delta \mathscr{R} / \delta Q=0$,
\begin{equation}
  \label{eq:Q0}
  Q = \frac{B}{12 \eta} \left( - \rho g + 2 \gamma \cos\theta \frac{ \partial (L'/A')}{\partial z} \right).
\end{equation}
Combined with the conservation equation (\ref{eq:cons}), we obtain the time evolution of the meniscus
\begin{equation}
  \label{eq:te0}
  \frac{\partial G}{\partial t} = \frac{1}{A'} \frac{\partial}{\partial z} 
  \left[ \frac{B}{12 \eta} \left( \rho g - 2 \gamma \cos\theta \frac{ \partial (L'/A')}{\partial z} \right) \right].
\end{equation}

\section{Power-law corner}

We consider an general corner formed by two surfaces which are power $n$ of $x$
\begin{equation}
  \label{eq:cornerN}
  E(x) = c x^n, \quad n \ge 1. 
\end{equation}
Note here the parameter $c$ has a dimension of [LENGTH]$^{-n+1}$. 
To derive the time evolution equation, we need the following  
\begin{eqnarray}
  B(G) &=& \frac{c^3}{3n+1} G^{3n+1} , \\
  A'(G) &=& c G^{n} , \\
  \frac{\partial (L'/A')}{\partial z} & \simeq & - \frac{n}{c G^{n+1}} \frac{\partial G}{\partial z} .
\end{eqnarray}
We have kept only terms of lowest order in $G$. 

The flux (\ref{eq:Q0}) becomes
\begin{equation}
  Q = \frac{c^3}{12 \eta (3n+1)} G^{3n+1} \left( - \rho g - \frac{ 2 n \gamma \cos\theta}{c}
    \frac{1}{G^{n+1}} \frac{\partial G}{\partial z} \right) .
\end{equation}

The time evolution equation (\ref{eq:te0}) becomes
\begin{equation}
  \frac{\partial G}{\partial t} = \frac{c^2}{12 \eta (3n+1) G^n} \frac{\partial}{\partial z} 
  \left[ G^{3n+1} \left( \rho g + \frac{ 2 n \gamma \cos\theta}{c}
    \frac{1}{G^{n+1}} \frac{\partial G}{\partial z} \right) \right] .
\end{equation}
Scaling the length and the time with the following constants
\begin{equation}
  H_c = \left( \frac{ 2 n \gamma \cos\theta }{ c \rho g} \right)^{1/(n+1)} , \quad
  t_c = \frac{ 12 \eta}{ c^2 \rho g H_c^{2n-1} } ,
\end{equation}
we convert the equation into a dimensionless form
\begin{eqnarray}
  \frac{\partial \tilde{G} }{\partial \tilde{t}} &=& \frac{1}{(3n+1) \tilde{G}^n} \frac{\partial}{\partial \tilde{z}} 
  \left[ \tilde{G}^{3n+1} \left(1 + \frac{1}{\tilde{G}^{n+1}} \frac{\partial \tilde{G}}{\partial \tilde{z}} \right) \right] \\
  \label{eq:te_n}
     &=& \tilde{G}^{2n} \frac{\partial \tilde{G}}{\partial \tilde{z}}
           + \frac{2n}{3n+1} \tilde{G}^{n-1} \left( \frac{\partial \tilde{G}}{\partial \tilde{z}} \right)^2 
           + \frac{1}{3n+1} \tilde{G}^n \frac{\partial^2 \tilde{G}}{\partial \tilde{z}^2} ,
\end{eqnarray}
where the tildes denote the corresponding dimensionless variables.
This is the generalization of the equation which \citet{Higuera2008} derived for the capillary rising in the corner made of flat planes.

The equilibrium profile of the meniscus is then given by 
\begin{equation}
  \label{eq:eq_n}
  1 + \frac{1}{\tilde{G}^{n+1}} \frac{\partial \tilde{G}}{\partial \tilde{z}} = 0 
  \quad \Rightarrow \quad
  \tilde{G} = (n \tilde{z})^{-1/n} .
\end{equation}
Note that the equilibrium profile is unbounded at the edge ($\tilde{z} \rightarrow \infty$ as $\tilde{G} \rightarrow 0$).

The time evolution equation (\ref{eq:te_n}) admits a self-similar solution of the form 
\begin{equation}
  \tilde{G}( \tilde{z}, \tilde{t}) = F(\chi) \tilde{t}^{\alpha}, \quad 
  \chi = \tilde{z} \tilde{t}^{\beta} ,
\end{equation}
where $\alpha$ and $\beta$ are parameters to be determined.
Using the above expressions, we rewrite equation (\ref{eq:te_n}) as
\begin{equation}
  \label{eq:ss_n2}
  ( \beta \chi F' + \alpha F) \, \tilde{t}^{\alpha -1}  
  = F^{2n} F' \,  \tilde{t}^{(2n+1)\alpha + \beta}  
  + \left( \frac{2n}{3n+1} F^{n-1} (F')^2 + \frac{1}{3n+1} F^n F'' \right)
  \tilde{t}^{(n+1)\alpha + 2\beta} .
\end{equation}
The above equation becomes time-independent if 
\begin{equation}
  \alpha - 1 = (2n+1) \alpha + \beta = (n+1) \alpha + 2 \beta ,
\end{equation}
which leads to 
\begin{equation}
  \alpha = - \frac{1}{3n} , \quad \beta = - \frac{1}{3}. 
\end{equation}
Equation (\ref{eq:ss_n2}) then becomes an ordinary differential equation
\begin{equation}
  \label{eq:ss_n}
  F^{2n} F' + \frac{2n}{3n+1} F^{n-1} (F')^2 + \frac{1}{3n+1} F^n F'' + \frac{1}{3} (\chi F' + \frac{1}{n} F) = 0 .
\end{equation}

The meniscus profile should converge to the equilibrium form $(n\tilde{z})^{-1/n}$ in the limit $\tilde{t} \rightarrow \infty$. 
This leads to the first boundary condition
\begin{equation}
  \label{eq:ssbc0_n}
  \chi \rightarrow 0, \quad F(\chi) \rightarrow (n \chi)^{-1/n} .
\end{equation}

The second boundary condition is that the profile $F(\chi)$ approaches zero at certain value $\chi=\chi_0$. 
Assuming $F(\chi)$ behaves like $(\chi_0 - \chi)^{\gamma}$ as $\chi \rightarrow \chi_{0-}$, each term in equation (\ref{eq:ss_n}) behaves like
\begin{eqnarray}
  F^{2n} F'   & \sim & (\chi_0 - \chi)^{(2n+1)\gamma-1} , \\
  F^{n-1} (F')^2 & \sim & (\chi_0 - \chi)^{(n+1)\gamma-2} , \\
  F^n F''  & \sim & (\chi_0 - \chi)^{(n+1)\gamma-2} , \\
  \chi F'  & \sim & \chi_0 (\chi_0 - \chi)^{\gamma-1} , \\
  F        & \sim & (\chi_0 - \chi)^{\gamma} . 
\end{eqnarray}
Anticipating $\gamma \le 1$, the dominating terms are $F^{n-1} (F')^2$, $F^n F''$, and $\chi F'$. 
Upon ignoring other terms, equation (\ref{eq:ss_n}) becomes
\begin{equation}
  \frac{2n}{3n+1} F^{n-1} (F')^2 + \frac{1}{3n+1} F^n F'' + \frac{1}{3} \chi_0 F'  = 0 .
\end{equation}
The solution to the above equation leads to the second boundary condition
\begin{equation}
  \label{eq:ssbc1_n}
  \chi \rightarrow \chi_0, \quad 
  F(\chi) \rightarrow \left[ \frac{n(3n+1)}{3(n+1)} \chi_0 (\chi_0 - \chi) \right]^{1/n} .
\end{equation}
Once the solution of (\ref{eq:ss_n}) with the two boundary conditions (\ref{eq:ssbc0_n}) and (\ref{eq:ssbc1_n}) is obtained, we get the asymptotic solution of the tip position 
\begin{equation}
  \tilde{Z}_m = \chi_0 \tilde{t}\, ^{1/3}.
\end{equation}
We can rewrite the above equation using the capillary length $a_c=\sqrt{\gamma/\rho g}$ as the length scale and $\eta a_c/\gamma$ as the time scale \citep{Ponomarenko2011}.
This leads to
\begin{equation}
  \label{eq:Zm_ac}
  \frac{Z_m}{a_c} = \chi_0 \left( \frac{ n^2 \cos^2\theta }{3} \right)^{1/3} 
    \left( \frac{ \gamma t}{\eta a_c} \right)^{1/3} .
\end{equation}
This result confirms the $t^{1/3}$ scaling proposed by \citet{Ponomarenko2011}.
Our theory also gives a prediction of the front factor 
\begin{equation}
  \label{eq:C}
  C = \chi_0 \left( \frac{n^2 \cos^2 \theta}{3} \right)^{1/3} ,
\end{equation}
which is in general dependent on the power of the corner (here $\chi_0$ is a function of $n$) and the contact angle $\theta$.
One interesting observation is that the $c$ parameter from equation (\ref{eq:cornerN}) does not appear in equation (\ref{eq:Zm_ac}), thus the detail of the corner does not affect the tip dynamics.

In the following, we shall study the numerical solutions for special cases of $n=1$ and 2,
which will be called {\em linear corner} and {\em quadratic corner}, respectively.

\section{Examples and discussion}

\subsection{linear corner}

As a first example, we study the classical case of corner formed by two flat planes.
For this case, the $E(x)$ function is given by
\begin{equation}
  \label{eq:corner1}
  E(x) = a x, \quad a \ll 1.
\end{equation}

The dimensionless form of the time evolution equation (\ref{eq:te_n}) is
\begin{equation}
  \label{eq:te_1}
  \frac{\partial \tilde{G}}{\partial \tilde{t}} = \frac{1}{4 \tilde{G}} \frac{\partial}{\partial \tilde{z}} 
  \left[ \tilde{G}^4 \left(1 + \frac{1}{\tilde{G}^2} \frac{\partial \tilde{G}}{\partial \tilde{z}} \right) \right] 
  = \tilde{G}^2 \frac{\partial \tilde{G}}{\partial \tilde{z}}
  + \frac{1}{2} \left( \frac{\partial \tilde{G}}{\partial \tilde{z}} \right)^2 
  + \frac{1}{4} \tilde{G} \frac{\partial^2 \tilde{G}}{\partial \tilde{z}^2}.
\end{equation}
This is consistent with \citet{Higuera2008} and our previous work \citep{2019_Taylor_rising}.

The equilibrium profile is given by 
\begin{equation}
  1 + \frac{1}{\tilde{G}^2} \frac{\partial \tilde{G}}{\partial \tilde{z}} = 0 
  \quad \Rightarrow \quad
  \tilde{G} = \frac{1}{\tilde{z}} .
\end{equation}

\begin{figure}
  \centering
  \includegraphics[width=0.48\textwidth]{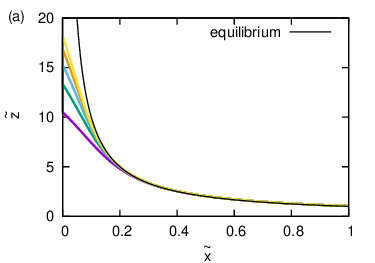}
  \includegraphics[width=0.48\textwidth]{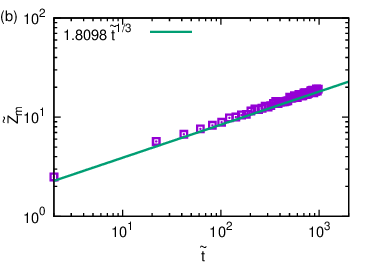} \\
  \includegraphics[width=0.48\textwidth]{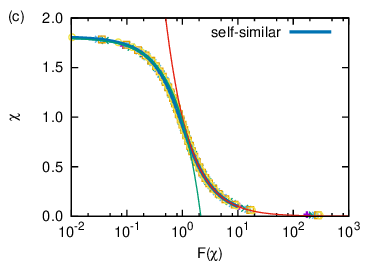}
  \includegraphics[width=0.48\textwidth]{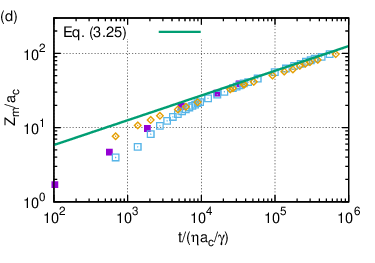}
  \caption{Linear corner: (a) Meniscus shape at different times ($\tilde{t}=$200, 400, 600, 800, 1000 from bottom to top). The solution are obtained by solving the time evolution equation (\ref{eq:te_1}). 
(b) The position of the tip as a function of time.
(c) Self-similar solution of equation (\ref{eq:ss_1}). Also shown are the meniscus shapes at different times (shown as symbols) and boundary conditions (\ref{eq:ssbc0_1}, red) and (\ref{eq:ssbc1_1}, green). (d) Comparison with experiments. Here the tip position is scaled by the capillary length $a_c$ and the time by $\eta a_c/\gamma$. 
$\blacksquare$ data from \citet{Higuera2008} ($a=2\tan 0.75^{\circ} \simeq 0.026$ and silicon oil V460); 
$\square$ data from \citet{Ponomarenko2011} ($a=2\tan 2.5^{\circ} \simeq 0.087$ and silicon oil V20); 
$\Diamond$ data from \citet{Ponomarenko2011} ($a=2\tan 6.5^{\circ} \simeq 0.228$ and silicon oil V20).} 
\label{fig:te_1}
\end{figure}

The time evolution equation (\ref{eq:te_1}) can be solved numerically (see Appendix for details). 
The meniscus profiles at different times are shown in figure~\ref{fig:te_1}(a).
The tip position as a function of time is shown in figure~\ref{fig:te_1}(b), which follows a $\tilde{t}^{\, 1/3}$ scaling.  

The self-similar solution has the form 
\begin{equation}
  \tilde{G}(\tilde{z}, \tilde{t}) = F(\chi) \tilde{t}^{-1/3}, \quad 
  \chi = \tilde{z} \tilde{t}^{-1/3} .
\end{equation}
The self-similar solution satisfies the equation (\ref{eq:ss_n}) with $n=1$
\begin{equation}
  \label{eq:ss_1}
  F^2 F' + \frac{1}{2} (F')^2 + \frac{1}{4} F F'' + \frac{1}{3} (\chi F)' = 0 ,
\end{equation}
and the following boundary conditions
\begin{eqnarray}
  \label{eq:ssbc0_1}
  && \chi \rightarrow 0, \quad F(\chi) \rightarrow 1/\chi , \\
  \label{eq:ssbc1_1}
  && \chi \rightarrow \chi_0, \quad F(\chi) \rightarrow \frac{2}{3} \chi_0 ( \chi_0 - \chi).
\end{eqnarray}
The numerical result of the self-similar solution is shown in figure~\ref{fig:te_1}(c), which gives
\begin{equation}
  \chi_0 \simeq 1.8098. 
\end{equation}
In figure~\ref{fig:te_1}(d), we also compare the prediction of equation (\ref{eq:Zm_ac}) to experimental data reported in \citet{Higuera2008} and \citet{Ponomarenko2011}, and good agreement is found at late times.

\subsection{quadratic corner}

We next examine the corner formed by two surfaces which are quadratic functions
\begin{equation}
  E(x) = b x^2. 
\end{equation}
Note here the parameter $b$ has a dimension of [LENGTH]$^{-1}$. 
The dimensionless form of the time evolution equations is
\begin{equation}
  \label{eq:te_2}
  \frac{\partial \tilde{G}}{\partial \tilde{t}} = \frac{1}{7 \tilde{G}^2} \frac{\partial}{\partial \tilde{z}} 
  \left[ \tilde{G}^7 \left(1 + \frac{1}{\tilde{G}^3} \frac{\partial \tilde{G}}{\partial \tilde{z}} \right) \right]
  = \tilde{G}^4 \frac{\partial \tilde{G}}{\partial \tilde{z}}
  + \frac{4}{7} \tilde{G} \left( \frac{\partial \tilde{G}}{\partial \tilde{z}} \right)^2 
  + \frac{1}{7} \tilde{G}^2 \frac{\partial^2 \tilde{G}}{\partial \tilde{z}^2}.
\end{equation}

The equilibrium profile is given by 
\begin{equation}
  1 + \frac{1}{\tilde{G}^3} \frac{\partial \tilde{G}}{\partial \tilde{z}} = 0 
  \quad \Rightarrow \quad
  \tilde{G} = (2 \tilde{z})^{-1/2} .
\end{equation}

\begin{figure}
  \centering
  \includegraphics[width=0.48\textwidth]{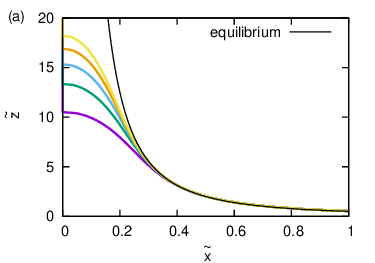}
  \includegraphics[width=0.48\textwidth]{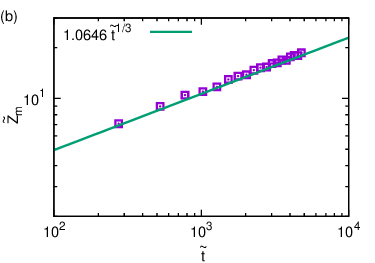} \\
  \includegraphics[width=0.48\textwidth]{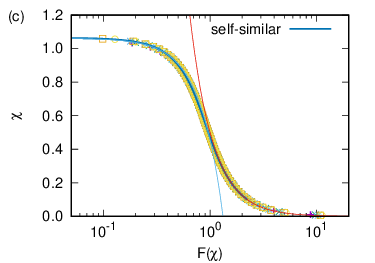}
  \includegraphics[width=0.48\textwidth]{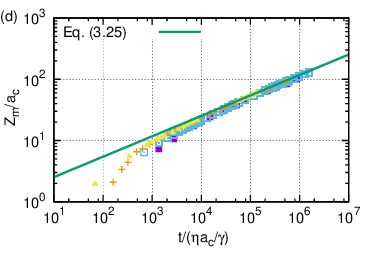}
  \caption{Quadratic corner: (a) Meniscus shape at different times ($\tilde{t}=$1000, 2000, 3000, 4000, 50000 from bottom to top). The solution are obtained by solving the time evolution equation (\ref{eq:te_2}). 
(b) The position of the tip as a function of time.
(c) Self-similar solution of equation (\ref{eq:ss_2}). Also shown are the meniscus shapes at different times (shown as symbols) and boundary conditions (\ref{eq:ssbc0_2}, red) and (\ref{eq:ssbc1_2}, green).
(d) Comparison with experiments from \citet{Ponomarenko2011}. Here the tip position is scaled by the capillary length $a_c$ and the time by $\eta a_c/\gamma$. 
The corner is given by $b=15$ mm$^{-1}$. 
$\blacksquare$ silicon oil V10;  
$\square$ silicon oil V20; 
$+$ silicon oil V170;
$\triangle$ silicon oil V1000.} 
  \label{fig:te_2}
\end{figure}

The time evolution of the profile is shown in figure \ref{fig:te_2}(a).
Near the tip, the shape of the meniscus for the quadratic corner is convex away from the corner edge, which is different to that for the linear corner.
The tip position as a function of time is shown in figure \ref{fig:te_2}(b), which also follows a $\tilde{t}^{\,1/3}$ scaling.  

The time evolution equation (\ref{eq:te_2}) admits a self-similar solution of the form 
\begin{equation}
  G(z,t) = F(\chi) t^{-1/6}, \quad 
  \chi = z t^{-1/3} .
\end{equation}
The solution satisfies the equation
\begin{equation}
  \label{eq:ss_2}
  F^4 F' + \frac{4}{7} F (F')^2 + \frac{1}{7} F^2 F'' + \frac{1}{3} (\chi F' + \frac{1}{2} F ) = 0 .
\end{equation}
The boundary conditions are
\begin{eqnarray}
  \label{eq:ssbc0_2}
  && \chi \rightarrow 0, \quad F(\chi) \rightarrow (2 \chi)^{-1/2} \\
  \label{eq:ssbc1_2}
  && \chi \rightarrow \chi_0, \quad 
  F(\chi) \rightarrow \left[ \frac{14}{9} \chi_0 (\chi_0 - \chi) \right]^{1/2} .
\end{eqnarray}

The self-similar solution is shown in figure~\ref{fig:te_2}(c), which gives
\begin{equation}
  \chi_0 \simeq 1.0646. 
\end{equation}
In figure~\ref{fig:te_2}(d), we compare the prediction (\ref{eq:Zm_ac}) to experimental data reported in \citet{Ponomarenko2011}, and again good agreement is found at late times.

\subsection{power-law corners}

Power-law corners with $n>2$ show similar results as the quadratic corner. 
In Table \ref{tab:chi0}, we list numerical value of $\chi_0$ for power-law corners up to $n=5$.

\begin{table}
  \begin{center}
\def~{\hphantom{0}}
  \begin{tabular}{p{1.5cm}cc}
    $n$  & $\quad\quad$ $\chi_0$ $\quad\quad$ & $\quad\quad$ $C = \chi_0 (n^2/3)^{1/3}$ $\quad\quad$ \\[3pt]
    1   & 1.8098 & 1.255 \\
    2   & 1.0646 & 1.172 \\
    3   & 0.7882 & 1.137 \\
    4   & 0.6395 & 1.117 \\
    5   & 0.6175 & 1.252 \\
  \end{tabular}
  \caption{Values of $\chi_0$ and $C=\chi_0(n^2/3)^{1/3}$ for power-law corners up to $n=5$.}
  \label{tab:chi0}
  \end{center}
\end{table}

\begin{figure}
  \centering
  \includegraphics[width=0.8\textwidth]{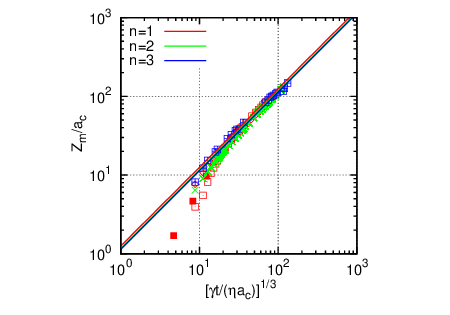}
  \caption{Comparison of numerical calculation and experimental results for different corners. 
Here the tip position is scaled by the capillary length $a_c$ and the $x$-axis is the scaled time to the power of $1/3$, $[\gamma t/(\eta a_c)]^{1/3}$.
Experimental data are from \citet{Higuera2008} and \citet{Ponomarenko2011}. 
(i) linear corners ($E=ax$): 
$\blacksquare$ $a=2\tan 0.75^{\circ} \simeq 0.026$ and silicon oil V460; 
$\square$ $a=2\tan 2.5^{\circ} \simeq 0.087$ and silicon oil V20;
$\Diamond$ $a=2\tan 6.5^{\circ} \simeq 0.228$ and silicon oil V20.
(ii) quadratic corner ($E=bx^2$): 
$\times$ $b=15 \textrm{ mm}^{-1}$ and silicon oil V20.
(iii) cubic corner ($E=cx^3$):
$\boxplus$ $c=18 \textrm{ cm}^{-2}$ and silicon oil V20.}
  \label{fig:te_3}
\end{figure}

\citet{Ponomarenko2011} suggested that the evolution of the meniscus front for \emph{different} corners would collapse if one scales the height with the capillary length $a_c$ and the time with $\eta a_c/\gamma$. 
In our formulation, this corresponds to the front factor $C$ of equation (\ref{eq:C}) is independent of $n$. 
This is a strong prediction.
In Table \ref{tab:chi0}, we also list the front factor $C$ for fully-wetting liquid ($\theta=0$). 
One interesting observation is that even the value of $\chi_0$ are different, but the front factors are similar with a numerical value around 1.2. 
Thus our numerical results support the proposition of \citet{Ponomarenko2011}. 
Figure~\ref{fig:te_3} shows the numerical results for linear, quadratic, and cubic corners, together with the experimental results from \citet{Higuera2008} and \citet{Ponomarenko2011}.

\subsection{discussion}
\label{sec:discussion}

Here we discuss the limitation of our model and various assumptions used in our calculation. 
From the equilibrium profile (\ref{eq:eq_n}), one sees that the meniscus is unbounded at the edge. 
This is not always the case (the circular tube is a typical counter-example) and depends on the corner geometry and the wettability of the fluid.
If the height of the equilibrium meniscus is finite, then our theory does not apply. 
For linear corners, the unbounded condition is given by \citet{Concus1969}, $\alpha + \theta < \pi/2$, where $\alpha$ is the open angle of the corner and $\theta$ is the equilibrium contact angle. 
Since we consider the wetting fluid with an equilibrium contact angle $\theta$ close to zero, the above condition is satisfied for the linear corner with a small open angle. 
For power-law corners, the open angle is zero as $\tilde{x} \rightarrow 0$ , so the condition is again satisfied.

Our analysis focused on to the long-time dynamics of the meniscus. 
The early stage of the capillary rising is complicated and various factors come into play. 
At early times when the meniscus height is less than the capillary length $a_c$, the effect of the gravity can be neglected. 
\citet{Quere1997} had shown the initial rising follows $t^1$ scaling when inertial is important.
Another complication arises during the onset of the meniscus. 
When the corner first touches the liquid, the assumption of equilibrium contact angle is not fulfilled and it takes about $10^2$--$10^3$ $\eta a_c/\gamma$ to establish the equilibrium contact angle \citep{Clanet2002}. 
In our study, the effect of inertial is neglected and we are working in the viscous-dominating region where the Reynolds number is close to zero.

When the inertial is neglected, the balance of capillary and viscous forces at early times leads to the classical Lucas-Washburn $t^{1/2}$ scaling. 
For linear corners, \citet{Higuera2008} had showed that the height of the meniscus at distance $\tilde{x}$ away from the edge evolves like $\tilde{H}(\tilde{x}) \sim \tilde{x}^{1/2} \, \tilde{t}^{\,1/2}$. 
For power-law corners, one can show $\tilde{H}(\tilde{x}) \sim \tilde{x}^{n/2} \, \tilde{t}^{\,1/2}$. 
The equilibrium height profile is $\tilde{H}_{\rm e}(\tilde{x}) \sim \tilde{x}^{-n}$, which can be obtained by inverting equation (\ref{eq:eq_n}). 
Equating these two heights leads to a characteristic time for the meniscus at $\tilde{x}$ to reach its equilibrium height, $\tilde{t}_{\rm e} \sim \tilde{x}^{-3n}$. 
Thus, the meniscus with larger $\tilde{x}$ equilibrates faster.
In this study, we focus on the asymptotic dynamics when the meniscus far from the edge is nearly at equilibrium while close to the edge the meniscus is still rising.

In writing the Rayleighian in terms of (\ref{eq:Ray}), we have made two assumptions regarding the shape of the meniscus. 
In the free energy part, we have neglected the energy contribution from the free surface. 
This corresponds to corners with small opening, $E(G) \ll L(G) \sim G$, i.e., the separation between two planes is much smaller than the distance to the edge. 
For linear corners, this requires $a \ll 1$ in equation (\ref{eq:corner1}). 
For power-law corners, this condition leads to $c G^{n-1} \ll 1$ and becomes progressively better satisfied when the tip is approached, $G \rightarrow 1$.

In the dissipation function we have used the lubrication approximation and assume the flow is one-dimensional.
This requires that the slenderness parameter, the ratio between the length scale in the $x$-direction to that in the $z$-direction \citep{Weislogel1998, Weislogel2001, Weislogel2011}, to be small. 
For the meniscus at position $\tilde{x}$, it takes about $\tilde{t}_{\rm e} \sim \tilde{x}^{-3n}$ to reach the equilibrium. 
At this time, the tip position is at $\tilde{Z}_m \sim \tilde{t}_{\rm e}^{\,1/3} \sim \tilde{x}^{-n}$.  
The slenderness parameter $\sim \tilde{x}/\tilde{Z}_m \sim \tilde{x}^{n+1}$, which is small when $\tilde{x}<1$ and become smaller when the rise proceeds.

\section{Conclusion}

We have studied the capillary rising of wetting fluid in small interior corners. 
For the general power-law corners with small opening angle, we used the Onsager variational principle to derive a time evolution equation for the meniscus profile.
The time evolution equation has a self-similar solution, and we have showed that the advance of the meniscus front follows a universal $t^{1/3}$ law. 
The universality of the $t^{1/3}$ scaling was previously demonstrated in experiments \citep{Ponomarenko2011}. 
Here we have shown that the $t^{1/3}$ scaling is indeed satisfied for general power-law corners. 
Furthermore, we have compute explicitly the prefactor, which only depends on the power $n$ and the equilibrium contact angle $\theta$.

This work was supported by the National Natural Science Foundation of China (NSFC) through the Grant No. 21774004. 
M.D. acknowledges the financial support of the Chinese Central Government in the Thousand Talents Program.
We thank Alexandre Ponomarenko, David Qu{\'e}r{\'e}, and Christophe Clanet for providing the experimental data in their paper \citep{Ponomarenko2011}.
We also would like to acknowledge the anonymous referee who raised the point on the prefactor.

Declaration of Interests. The authors report no conflict of interest.

\appendix
\section{Numerical solution to equation (\ref{eq:te_n})}
\label{sec:app}

The time evolution equation of the meniscus profile is given by equation (\ref{eq:te_n}).
For $n=1$ and $n=2$, this equation takes the form of (\ref{eq:te_1}) and (\ref{eq:te_2}), respectively.
These partial differential equations can be solved numerically using Matlab.

The range of $\tilde{z}$ is $[0.1:20]$.
The lower boundary condition is given by the equilibrium profile 
\begin{equation}
  \tilde{G}(\tilde{z}=0.1)=(n \times 0.1)^{-1/n}, 
\end{equation}
and we assume the part of the meniscus far away from the edge ($\tilde{z}<0.1$) has already reached the equilibrium. 
The upper boundary is chosen to be large enough so the tip position does not exceed the upper boundary at the end of calculation. 
In this study, we used 
\begin{equation}
  \tilde{G}(\tilde{z}=20)=0.
\end{equation}

The initial condition is a straight line connecting the lower boundary and the meniscus tip $\tilde{Z}_{m0}$, where $\tilde{Z}_{m0}$ is the initial tip position and takes a small value. 
Different choices of $\tilde{Z}_{m0}$ do not affect the long-time dynamics. 
The results shown in figures \ref{fig:te_1} and \ref{fig:te_2} are obtained by setting $\tilde{Z}_{m0}=1.0$.


\bibliographystyle{jfm}
\bibliography{Taylor}

@ARTICLE{Clanet2002,
  author = {Christophe Clanet and David Qu{\'e}r{\'e}},
  title = {Onset of menisci},
  journal = {J. Fluid Mech.},
  year = {2002},
  volume = {460},
  pages = {131--149},
  doi = {10.1017/s002211200200808x},
  publisher = {Cambridge University Press ({CUP})}
}

@ARTICLE{Concus1969,
  author = {Paul Concus and Robert Finn},
  title = {On the behavior of a capillary surface in a wedge},
  journal = {Proc. Natl. Acad. Sci. U.S.A.},
  year = {1969},
  volume = {63},
  pages = {292--299}
}

@BOOK{DoiSoft,
  title = {Soft Matter Physics},
  publisher = {Oxford University Press},
  year = {2013},
  author = {Masao Doi}
}

@ARTICLE{Dong1995,
  author = {M. Dong and I. Chatzis},
  title = {The Imbibition and Flow of a Wetting Liquid along the Corners of
	a Square Capillary Tube},
  journal = {J. Colloid Interface Sci.},
  year = {1995},
  volume = {172},
  pages = {278--288},
  month = {jun},
  doi = {10.1006/jcis.1995.1253},
  publisher = {Elsevier {BV}}
}

@ARTICLE{Finn2002a,
  author = {Robert Finn},
  title = {Some Properties of Capillary Surfaces},
  journal = {Milan Journal of Mathematics},
  year = {2002},
  volume = {70},
  pages = {1--23},
  month = {sep},
  doi = {10.1007/s00032-002-0001-y},
  publisher = {Springer Nature}
}

@ARTICLE{Finn1999,
  author = {Robert Finn},
  title = {Capillary Surface Interfaces},
  journal = {Notices of AMS},
  year = {1999},
  volume = {46},
  pages = {770--781}
}

@BOOK{dBQ,
  title = {Capillarity and Wetting Phenomena},
  publisher = {Springer},
  year = {2004},
  author = {de Gennes, Pierre-Gilles and Brochard-Wyart, Fran{\c c}oise and Qu{\'e}r{\'e},
	David}
}

@ARTICLE{Hauksbee1710,
  author = {Francis Hauksbee},
  title = {X. {An} account of an experiment touching the ascent of water between
	two glass planes, in an hyperbolick figure},
  journal = {Philos. Trans. R. Soc. London},
  year = {1710},
  volume = {27},
  pages = {539--540},
  month = {jan},
  doi = {10.1098/rstl.1710.0071},
  publisher = {The Royal Society}
}

@ARTICLE{Higuera2008,
  author = {F. J. Higuera and A. Medina and A. Li{\~{n}}{\'{a}}n},
  title = {Capillary rise of a liquid between two vertical plates making a small
	angle},
  journal = {Phys. Fluids},
  year = {2008},
  volume = {20},
  pages = {102102},
  month = {oct},
  doi = {10.1063/1.3000425},
  publisher = {{AIP} Publishing}
}

@ARTICLE{Langbein1990,
  author = {Dieter Langbein},
  title = {The shape and stability of liquid menisci at solid edges},
  journal = {J. Fluid Mech.},
  year = {1990},
  volume = {213},
  pages = {251--265},
  month = {apr},
  doi = {10.1017/s0022112090002312},
  publisher = {Cambridge University Press ({CUP})}
}

@ARTICLE{Lucas1918,
  author = {R. Lucas},
  title = {Ueber das Zeitgesetz des kapillaren Aufstiegs von Fl{\"u}ssigkeiten},
  journal = {Kolloid-Zeitschrift},
  year = {1918},
  volume = {23},
  pages = {15--22},
  doi = {10.1007/BF01461107}
}

@ARTICLE{Onsager1931,
  author = {Lars Onsager},
  title = {Reciprocal Relations in Irreversible Processes. {I}.},
  journal = {Phys. Rev.},
  year = {1931},
  volume = {37},
  pages = {405--426},
  month = {feb},
  doi = {10.1103/physrev.37.405},
  publisher = {American Physical Society ({APS})}
}

@ARTICLE{Onsager1931a,
  author = {Lars Onsager},
  title = {Reciprocal Relations in Irreversible Processes. {II}.},
  journal = {Phys. Rev.},
  year = {1931},
  volume = {38},
  pages = {2265--2279},
  month = {dec},
  doi = {10.1103/physrev.38.2265},
  publisher = {American Physical Society ({APS})}
}

@ARTICLE{Ponomarenko2011,
  author = {Alexandre Ponomarenko and David Qu{\'e}r{\'e} and Christophe Clanet},
  title = {A universal law for capillary rise in corners},
  journal = {J. Fluid Mech.},
  year = {2011},
  volume = {666},
  pages = {146--154},
  month = {jan},
  doi = {10.1017/s0022112010005276},
  publisher = {Cambridge University Press ({CUP})}
}

@ARTICLE{Quere1997,
  author = {David Qu{\'{e}}r{\'{e}}},
  title = {Inertial capillarity},
  journal = {Europhys. Lett.},
  year = {1997},
  volume = {39},
  pages = {533--538},
  doi = {10.1209/epl/i1997-00389-2},
  publisher = {{IOP} Publishing}
}

@ARTICLE{Tang1994,
  author = {Lei-Han Tang and Yu Tang},
  title = {Capillary rise in tubes with sharp grooves},
  journal = {J. Phys. {II}},
  year = {1994},
  volume = {4},
  pages = {881--890},
  month = {may},
  doi = {10.1051/jp2:1994172},
  publisher = {{EDP} Sciences}
}

@ARTICLE{Taylor1710,
  author = {Brook Taylor},
  title = {{IX}. {Part} of a letter from {Mr. Brook Taylor, F. R. S. to Dr.
	Hans Sloane R. S. Secr.} concerning the ascent of water between two
	glass planes},
  journal = {Phil. Trans. R. Soc. London},
  year = {1710},
  volume = {27},
  pages = {538--538},
  month = {jan},
  doi = {10.1098/rstl.1710.0070},
  publisher = {The Royal Society}
}

@ARTICLE{Washburn1921,
  author = {Edward W. Washburn},
  title = {The Dynamics of Capillary Flow},
  journal = {Phys. Rev.},
  year = {1921},
  volume = {17},
  pages = {273--283},
  month = {mar},
  doi = {10.1103/physrev.17.273},
  publisher = {American Physical Society ({APS})}
}

@ARTICLE{Weislogel2001,
  author = {Mark M. Weislogel},
  title = {Capillary flow in interior corners: The infinite column},
  journal = {Phys. Fluids},
  year = {2001},
  volume = {13},
  pages = {3101},
  doi = {10.1063/1.1408918},
  publisher = {{AIP} Publishing}
}

@ARTICLE{Weislogel2011,
  author = {Mark M. Weislogel and J. Alex Baker and Ryan M. Jenson},
  title = {Quasi-steady capillarity-driven flows in slender containers with
	interior edges},
  journal = {J. Fluid Mech.},
  year = {2011},
  volume = {685},
  pages = {271--305},
  month = {sep},
  doi = {10.1017/jfm.2011.314},
  publisher = {Cambridge University Press ({CUP})}
}

@ARTICLE{Weislogel1998,
  author = {Mark M. Weislogel and Seth Lichter},
  title = {Capillary flow in an interior corner},
  journal = {J. Fluid Mech.},
  year = {1998},
  volume = {373},
  pages = {349--378},
  month = {oct},
  doi = {10.1017/s0022112098002535},
  publisher = {Cambridge University Press ({CUP})}
}

@ARTICLE{2019_Taylor_rising,
  author = {Tian Yu and Ying Jiang and Jiajia Zhou and Masao Doi},
  title = {Dynamics of {Taylor} rising},
  journal = {Langmuir},
  year = {2019},
  volume = {35},
  pages = {5183--5190},
  doi = {10.1021/acs.langmuir.9b00335},
  url = {https://arxiv.org/abs/1903.11522}
}

\end{document}